\documentclass[
    ,final            
  ]
  {aipproc}

\layoutstyle{6x9}

\def\pt{p_T}

\begin{document}

\title{Prospects for (non-SUSY) new physics with first LHC data}


\author{Jonathan Butterworth, on behalf of the ATLAS and CMS collaborations}{
  address={Department of Physics and Astronomy, University College London, 
  Gower St. WC1E 6BT, UK}
}

\begin{abstract}
The ATLAS and CMS experiments will take first data soon. I
consider here the prospects for new physics (excluding SUSY) with a
few fb$^{-1}$ of data. This means processes with signal cross sections
of a few 10 fb or more, with clear and fairly simple signatures -
precision comparison of data to Standard Model tails will take longer,
needing more luminosity and very good understanding of detector
calibrations, resolutions and trigger efficiencies. The approach I
take here is signature rather than model based, but examples of models
will be given.
\end{abstract}

\maketitle


\section{Introduction}

The Large Hadron Collider (LHC) will provide proton-proton collisions
at 14 TeV - a factor of seven above previous colliders. This is not
just ``any old'' new energy regime either. It will give us the first
direct look at physics far above the electroweak-symmetry-breaking
scale.  Thus if the Standard Model (SM) is treated as a Higgsless
low-energy effective theory, calculations of SM cross sections violate
unitarity within the experimental reach of the experiments. Thus the
experiments must see either the SM Higgs, or other new physics, or
both. This underlines the need for the experiments to be alert to the
widest range of possible exotic signatures. In this brief review I
discuss prospects for observing some of these signatures with early
data, and to which models they are sensitive.

\section{Various Resonances}

\begin{figure}
  \includegraphics[height=.3\textheight]{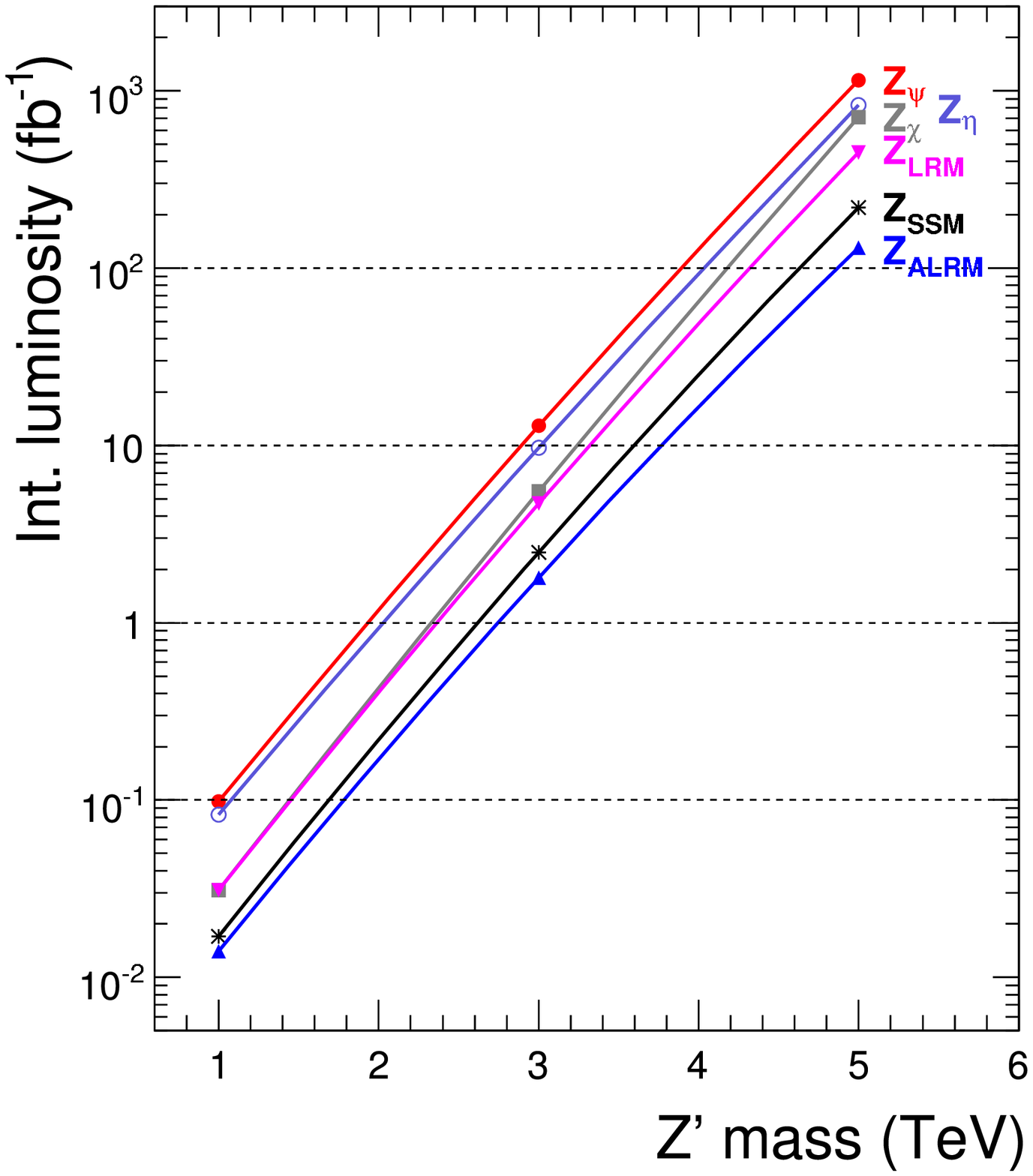}
  \includegraphics[height=.3\textheight]{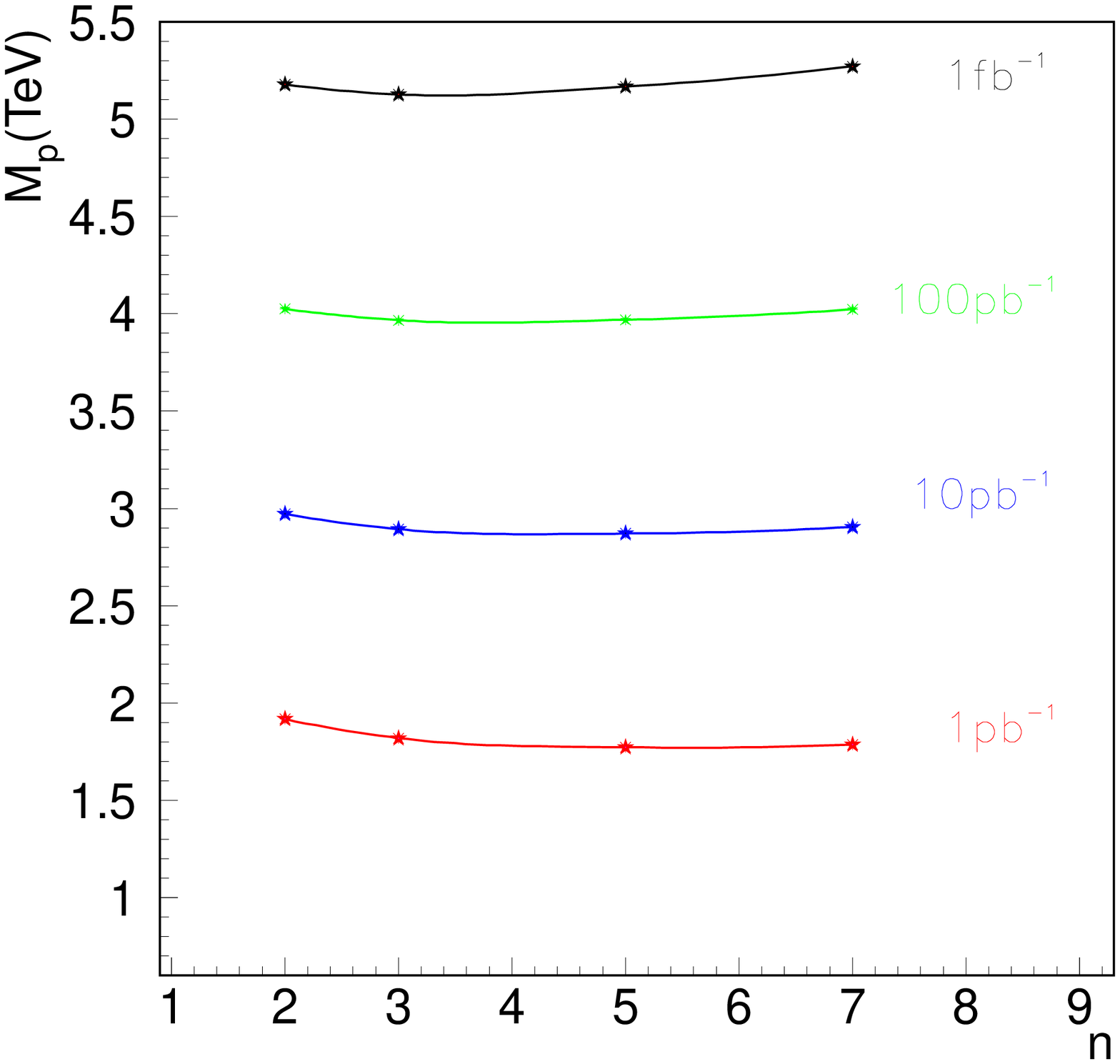}
  \caption{(l) CMS 5$\sigma$ discovery reach as a function of
  integrated luminosity as a function of $Z^\prime$ mass for various
  models with $Z^\prime \rightarrow \mu^+\mu^-$ (from \cite{cmstdr})
  (r) ATLAS discovery contours in the plane of Planck mass ($M_P$) and
  number of extra dimensions ($n$) for black hole masses more than
  1~TeV above $M_P$ (from~\cite{Tanaka:2004xb})}
  \label{fig:fig1}
\end{figure}

One of the simplest places to start is looking for anomalous, in
particular resonant, production of dileptons. The CMS collaboration
has recently~\cite{cmstdr} presented a sensitivity study of
$\mu^+\mu^-$ resonances above the Drell-Yan background. Such signals
can arise for example from $Z^\prime$ bosons present in grand
unification models~\cite{Leike:1998wr}. With as little as 0.1
fb$^{-1}$ there is sensitivity to a 1~TeV $Z^\prime$ in all the models
considered, even when a realistic estimate of the detector alignment
achievable with this luminosity is included.  With 1~fb$^{-1}$ there
is still sensitivity up to around 2.5 TeV, dependent upon the specific
model - see Fig.~\ref{fig:fig1} (left). Such studies also give an
indication of sensitivity to graviton resonances in Randall-Sundrum
models, for example. One might begin to distinguish spin-1 from spin-2
resonances with ~50 fb$^{-1}$.

Graviton resonances are an example where one also expects to see
resonances in the diphoton channel. Here the CMS experiment
expects~\cite{cmstdr,Lemaire:2006kf} to have sensitivity up to around
3.5~TeV with 10~fb$^{-1}$.

Kaluza-Klein excitations of the gluon in TeV$^{-1}$ extra dimension
models are an example of a model producing resonances decaying to
quarks. ATLAS has carried out a study~\cite{ros} of 1~TeV resonances
decaying to $t\bar{t}$, where the $W$ from one of the tops decays to
hadrons, the other to (e or $\mu$) + $\nu$.  There is sensitivity here
with around a fb$^{-1}$, and the mass reach eventually is expected to be
up to around 3.3~TeV.

\section{Missing Momentum}

It is natural to benchmark signatures with missing transverse momentum
plus leptons against models with high mass $W^\prime$ bosons, just as the
dilepton case is benchmarked against $Z^\prime$ models. Discovery
limits are typically based on a measured transverse mass distribution,
after a selection requiring the presence of a high transverse momentum
($\pt$) lepton. An example study from CMS~\cite{cmstdr,Hof:2006bu}
using a reference model~\cite{Altarelli} shows sensitivity up to
around 3.5 TeV in the $W^\prime$ mass with 1~fb$^{-1}$ of data.

Multiple lepton signatures in association with missing $\pt$ arise
naturally in models where a complex decay chain ensues from a massive
particle decay. For example in universal extra dimension
models~\cite{Appelquist:2000nn}, Kaluza-Klein excitations of the gluon
occur, carrying a conserved quantum number which plays a similar role
to R-parity in the minimal supersymmetric SM. That is, decays follow
through a cascade (which may produce leptons) ending in the lightest
Kaluza-Klein state, which cannot decay and carries off ``missing''
momentum. A study from CMS~\cite{Kazana:2007zz} shows sensitivity
after a few fb$^{-1}$ of up to around 700~GeV in the mass scale
associated with the size of the extra dimensions in the minimal
universal extra dimension model.

\section{Jets and Vector Bosons}

Jets themselves will give the highest statistics probe of the highest
energy scales at the LHC (unless they are suppressed by other exotic
physics such as TeV black holes). The major limitation on the
sensitivity is likely to be from systematic errors at first associated
with detector, where understanding the energy measurement for jets of
several TeV will require much detailed study. In the end, as the
detectors become better understood, theoretical uncertainties on the
QCD cross section will also play a role. That said, both these effects
can be reduced by considering the ratio of jets in different rapidity
regions. In this case, a CMS study~\cite{cmstdr} indicates that the
dijet mass ratios could be measured up to 6.5 TeV with an accuracy of
better than 50\%, giving sensitivity to new physics above 15 TeV if
the effects are manifested as effective contact interactions.

A qualitatively new feature expected at the LHC is the abundance of
particles with mass {\cal O}(100) GeV and $\pt$ of several hundred GeV
decaying to hadrons - SM candidates being the $W$, $Z$, top and Higgs,
but with other exotica also possible.  The boost means that the decay
products will often appear in a single jet.  Thus measuring the
single-jet mass, and studying substructure of the jet for signs of the
heavy particle decay, provides information which can be exploited in
the search for and diagnosis of new physics. This was first studied in
the case of vector boson scattering at high masses~\cite{bcf} and has
since been proposed and/or studied in several other
cases~\cite{Holdom:2007nw,Skiba:2007fw,Lillie:2007yh,Fleming:2007qr,Baur:2007ck,Butterworth:2007ke},
including new physics searches ranging from superheavy quarks to
Kaluza-Klein graviton excitations, as well as supersymmetric decay
chains and SM measurements. Preliminary studies from ATLAS show that
single-jet mass resolutions of a few GeV should be obtainable, which
is good enough to make the method useful. For example, $t^\prime$
candidates may be observable using this technique with only
2.5~fb$^{-1}$ of data~\cite{Holdom:2007nw}.

Vector boson scattering, where a $W$ or $Z$ is radiated from a quark
in each proton and they then interact, is a process of great interest
over the whole range of centre-of-mass energy accessible at the
LHC. It is characterised by the $W/Z$ decay products and two high
rapidity ``tag'' jets travelling close to the beam directions. At low
masses, $WW$ and $ZZ$ scattering are search channels for the SM
Higgs. At high masses, unitarity is violated in this process without a
Higgs, so it represents the ``no lose'' channel at the LHC. If a
Higgs is seen, measuring this process is vital to show that the Higgs
mechanism actually operates. If no Higgs is seen, some other new
physics must appear in this channel to regulate the cross section at
the highest masses.

Of course, in its extreme this is a very high luminosity topic, where
the kinematic limit is probed only with very high luminosity (a
luminosity upgrade is in fact required to exploit it fully). However,
diboson resonances in this channel are a generic feature of models
involving new strong interactions around 1~TeV~\cite{Hill:2002ap}. For
example, studies by CMS and at ATLAS of $WZ$ and $WW$ resonances
respectively~\cite{cmstdr,Kreuzer:2006sd,stat} indicate that first
sensitivity will be obtained after a few fb$^{-1}$.

\section{Multiple jets and leptons}

In extra dimension models, the Planck scale may be lowered far enough
to be accessible at the LHC. This would lead to copious production of
mini black-holes, which would decay very rapidly via Hawking radiation
to ``flavour democratic'' high multiplicity states. Cuts on
multiplicity and event shapes (such as circularity) can be used to
extract as signal. There are large uncertainties in the phenomenology,
but the cross sections may be enormous, and there are cases where this
physics would be visible with very low
luminosities~\cite{Tanaka:2004xb}, down even to 1~pb$^{-1}$, once the
detector and backgrounds are sufficiently well understood! - see
Fig.\ref{fig:fig1} (right).

\section{Summary}

Many surprises are possible in the early days of LHC running.  A
``bonfire of models'' is certain, but some (perhaps only one?) will
survive. I have not discussed at all the enormous complexity of the
detectors and the huge challenges involved in understanding them, and
in understanding the SM at 14 TeV. The first things we see
will be features of new detectors - so buyer beware rumours and blogs!
Nevertheless, there will be some really exciting physics.


\begin{theacknowledgments}
Thanks especially to G. Brooijmans, S. Eno and F. Ledroit, as well as
to the organisers for a stimulating and fun conference.
\end{theacknowledgments}



\bibliographystyle{aipproc}   

\bibliography{jmb}

\IfFileExists{\jobname.bbl}{}
 {\typeout{}
  \typeout{******************************************}
  \typeout{** Please run "bibtex \jobname" to optain}
  \typeout{** the bibliography and then re-run LaTeX}
  \typeout{** twice to fix the references!}
  \typeout{******************************************}
  \typeout{}
 }

\end{document}